\documentclass[10pt,conference]{IEEEtran}

\IEEEoverridecommandlockouts

\usepackage[cmex10]{amsmath}
\usepackage{textcomp}
\usepackage{amssymb}

\usepackage{cite}
\usepackage{graphicx}
\usepackage{color}

\usepackage{mathtools}

\newtheorem{theorem}{Theorem}
\newtheorem{lemma}{Lemma}

\newcommand{\Xc}{\mathcal{X}}

\begin{document}

% paper title
\title{Joint Transfer of Energy and Information in a Two-hop Relay Channel} %
%\vspace{0ex}
\author{\normalsize Ali H. Abdollahi Bafghi, Mahtab~Mirmohseni, and~Mohammad Reza Aref\\

Information Systems and Security Lab (ISSL)\\

Department of Electrical Engineering, Sharif University of Technology, Tehran, Iran\\

Email: { hajiabdollahi\_a}@ee.sharif.edu,\{mirmohseni,aref\}@sharif.edu \vspace{0ex}
\thanks{This work was supported in part by the Iran National Science Foundation under Grant 92-32575.}}

%\markboth{}%

\maketitle

\begin{abstract}
We study the problem of joint information and energy transfer in a two-hop channel with a Radio frequency (RF) energy harvesting relay. We consider a finite battery size at the relay and deterministic energy loss in transmitting energy. In other words, to be able to send an energy-contained symbol, the relay must receive multiple energy-contained symbols. Thus, we face a kind of channel with memory. We model the energy saved in battery as channel state with the challenge that the receiver does not know the channel state. First, we consider the problem without any channel noise and derive an achievable rate. Next, we extend the results to the case with an independent and identically distributed noise in the second hop (the relay-receiver link).
\end{abstract}

\section{Introduction}
Nowadays energy consumption becomes an important design factor in communication systems instead of traditional parameters like throughput because of financial reasons and environmental concerns. There would be many equipments with large energy consumption in next generation (5G) networks, so energy efficiency will play an important rule in these networks \cite{Buzzi}.

One of alternative techniques used for energy consumption management is energy harvesting. Energy harvesting enables wireless networks to use environmental energies to increase energy efficiency, so required energy for communication nodes in wireless networks decrease noticeably.
This promising method is introduced in two main directions: (i) in the first direction, energy is harvested from environmental sources like wind and sunlight. The main characteristic of this setup is the sporadic nature of the harvested energy which makes the exact analysis rather difficult; (ii) in the second direction, known as Radio frequency (RF) energy harvesting, energy is harvested from the radio waves in space. This technique is promising with an increasing demand profile and some commercialized products \cite{lu2015wireless}.
Due to less randomness in obtaining the harvested energy, the analysis and strategies could be simpler compared with the harvesting from the environment.

The RF energy can be transferred concurrently with the information signal in a wireless system, proposed as simultaneous wireless information and power transfer (SWIPT) \cite{lu2015wireless,Buzzi}. A useful scenario in this case is to use received energy for future transmissions. In this scenario, the design of encoder and decoder is a new open problem, because of the memory now appeared in the system. In addition, the existing works show an inherent trade-off in transmitting energy and information simultaneously \cite{lu2015wireless}. Thus to understand the interplay among energy and information and also to obtain the optimal coding structure in this scenario, an information theoretic model works.
Because of complexity of models (especially channels with memory), fundamental limits in this system are not noticed widely. In particular, it is not clear what are the properties of an optimal code in this system(the code which transmits information and energy simultaneously).

One of the encoding methods for the joint energy and information transfer uses a finite state Markov source to generate codewords \cite{Fouladgar}. The energy reception constraint can be modeled with a channel with constraint on output, whose capacity was derived by Gastpar in \cite{Gastpar}. He also extended this result to multiple access and Gaussian relay channels \cite{Gastpar}. The problem of transmission of optimal rate with constraint on minimum received energy was studied by Varshney in \cite{Varshney}, where the capacity-energy function was introduced and some of its features were characterized. A channel with stochastic power restriction was studied by Ozel and Ulukus \cite{Ozel}. They showed that the capacity of the AWGN channel with random power available at the transmitter is the same as the capacity of an AWGN channel with an average power constraint equal to the average recharge rate. Capacity of a multiple access channel with constraint on received power was derived by Fouladgar and Simeone \cite{Simeone}. They also studied the multi-hop channel with energy harvesting relay and introduced capacity-energy function. However, they did not consider a finite battery at the relay \cite{Simeone}.
The capacity of a point to point channel with an energy harvesting transmitter was determined by Tutuncuoglu et. al. \cite{ulukus}, where the battery size was assumed to be one.

The problem of interactive communication with energy transfer (re-transmit the received energy) with finite battery sizes (i.e., finite energy units available in the system) was studied by Popovski et. al. \cite{popovski}. They derived the inner and outer bound for a two way orthogonal channel with finite energy units. In their system model, two nodes have a constant sum of energy units. If a node wants to send symbol "1", it has to cost one energy unit, but sending symbol "0" does not cost any energy unit. Similarly if a node receives symbol "1", it can save one energy unit and receiving symbol "0" does not have any energy unit \cite{popovski}. One of the important challenges which is not considered in \cite{popovski} is that the receiver node cannot save energy of received signal entirely and an energy loss occurs.
This energy loss can be modeled by assuming that the receiver has to receive $m$ energy-contained symbols for sending a symbol with energy.
Another interactive scenario that can be studied to understand the nature of the optimal codes in a joint information and energy transfer is relay channel.

In this paper, we consider a two-hop channel with an RF energy harvesting relay, where the transmitter jointly transfers information and energy to the relay. The harvested energy at the relay is used to re-transmit the data to the receiver. We assume finite battery size at the relay. The energy loss in transmitting energy is modeled with a fixed deterministic reduction in energy. In fact, the relay must receive multiple energy-contained symbols to be able to send one energy-contained symbol. These limitations at the relay turn the problem to the transmission over a channel \emph{with states}, where the state shows the energy level at the relay's battery.
Hence, we face a kind of channel with memory.
Thus, the main questions are which rates would be achievable in these models and what the structure of the coding techniques are that achieve those rates. One of the main challenge in the code design is to make the receiver be able to decode the message without knowing the sequence of states.
We model the energy stored in relay's battery as channel state with the challenge that the receiver does not know the channel state. First, we consider the problem without any channel noise and derive an achievable rate. We propose a new block Markov coding based achievability scheme in which the random codebooks are generated for each state. We show that the received codewords in the receiver form Markov sources. We use this property and Asymptotic Equipartition Property (A.E.P) Theorem \cite{ash} to propose a new decoding strategy which does not need state sequence at the receiver.
Next, we consider the problem with an independent and identically distributed noise in the second hop (the relay-receiver link) and find a coding scheme which works in the noisy condition.
Thus, we extend our achievable results to the noisy case, where we modify the decoding strategy at the receiver by using typical decoding with the capability of decoding without knowing state sequence.

\section{System Model}
%{\color{red} write a notation part here. capital letters show ...}

\begin{figure}
\centering
\includegraphics[width=8cm]{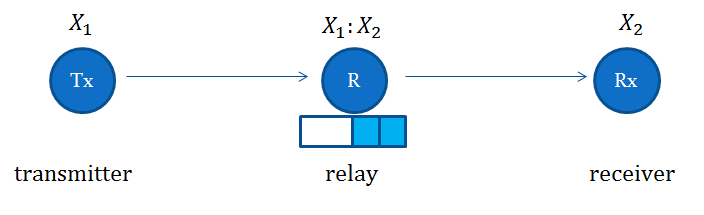}
\caption{Noiseless two-hop relay channel with finite battery (Noiseless THRC-FB)}
 \label{fig:noiseless}
\end{figure}

We consider a binary and noiseless two-hop relay channel illustrated in Fig. \ref{fig:noiseless}, in which the relay node has energy restriction (i.e., the relay must harvest energy from its received signal to be able to transmit). Also, we assume a finite battery at the relay which can save finite number of energy units. Thus, the transmitted symbol depends on the harvested energy from received symbols.

\textbf{Notation}: Upper-case letters (e.g., $X$) denote Random Variables (RVs) and lower-case letters (e.g., $x$) their realizations. The probability mass function (p.m.f) of a RV $X$ with alphabet set $\Xc$ is denoted by $p_X(x)$; occasionally, subscript $X$ is omitted.
%$A_\epsilon^n(X,Y)$ is the set of $\epsilon$-strongly, jointly typical sequences of length $n$.
$X^j_i$ indicates a sequence of RVs $(X_i,X_{i+1},...,X_j)$; we use $X^j$ instead of $X^j_1$ for brevity.
The channel inputs at the transmitter and the relay are shown by $\{ {X_{1,1}},{X_{1,2}},{X_{1,3}},...\}$ and $\{ {X_{2,1}},{X_{2,2}},{X_{2,3}},...\}$, respectively. The output at the relay  and the receiver's are denoted by $\{ {Y_{2,1}},{Y_{2,2}},{Y_{2,3}},...\}$ and $\{ {Y_{3,1}},{Y_{3,2}},{Y_{3,3}},...\}$, respectively. The channel input at the transmitter and the channel outputs at the relay and the receiver have binary alphabets, i.e., $\mathcal{X}_1=\mathcal{Y}_2=\mathcal{Y}_3=\{0,1\}$. The alphabet of channel input at the relay is shown by $\mathcal{X}_2$ (will be introduced later). $\{ {U_1},{U_2},{U_3},...\} $ show the level of battery storage in the $i$-th transmission which are used to model the channel state. $S_i$ are considered as: $S_1=(U_1,U_2),S_2=(U_2,U_3),S_3=(U_3,U_4),\ldots$ which are used in further proofs. $\pi_u$ denotes the steady state probability of the $u$-th state of the Markov chain.

In our system model transmitting symbol "1" needs $m$ energy units, while symbol "0" can be sent with no energy (the transmitting symbol "1" costs $m$ energy units for the relay). However, receiving symbol "1" at the relay charges the battery with only one energy unit. This consideration shows the energy loss in the channel.
The transmitter does not have energy restriction and it can transmit any symbol in each channel use. The transmitter sends a message $M\in [1,{{2}^{nR}}]$ to the relay node in $n$ channel uses (by transmitting $X_1^n$). Then, the relay decodes this message and sends it to the receiver (by transmitting $X_2^n$).
The $\mathcal{X}_{2,u}$ is the set of symbols which can be transmitted by relay when channel state is $u$. The channel state $u$ shows the energy units stored in the relay's battery and $U$ is the maximum battery size.
Energy restriction in the relay node is described as:
\begin{equation}
u < m \to {\mathcal{X} _{2,u}} \in \{ 0\}
\label{e1}
\end{equation}
\begin{equation}
u\ge m\to {{\mathcal{X} }_{2,u}}\in \{0,1\}
\label{e2}
\end{equation}
where $u \in [0:U]$. We call this system as noiseless two-hop relay channel with finite battery (Noiseless THRC-FB). Because of the noiseless channel property, the outputs are equal to inputs in each hop, i.e., $Y_2=X_1,Y_3=X_2$, where $Y_{2},Y_{3}$ are the received symbols at the relay and the receiver.

\begin{figure}
\centering
\includegraphics[width=7cm]{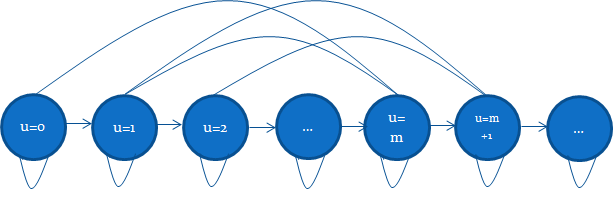}
\caption{State diagram of noiseless THRC-FB}
 \label{fig:state}
\end{figure}

The main difficulty here is that the system has memory due to the energy restriction and finite battery size at the relay. Our approach is to model the energy units (in the relay's battery) as the state of the system. Also, we assume that the receiver does not know the state sequence, which is another difficulty we face.
The state diagram of the channel is shown in Fig. \ref{fig:state}. As seen in Fig. \ref{fig:state}, when the battery is in state $u$ in the current transition, the following cases occur in the state diagram: i)If the relay receives symbol $1$ and transmits symbol $0$, the state in the next transition would be $u+1$ (except when $u=U$, where the next state won't change); ii)If the relay receives symbol $0$ and transmits symbol $1$, the state in the next transition would be $u-m$ (in this case we must have $u>m-1$, otherwise it does not occur); iii)If the relay receives symbol $1$ and transmits symbol $1$, the state in the next transition would be $u-m+1$ (in this case we must have $u>m-2$, otherwise it does not occur); iv)If the relay receives symbol $0$ and transmits symbol $0$, the next state would be the same as the current state.

\begin{figure}
\centering
\includegraphics[width=7.5cm]{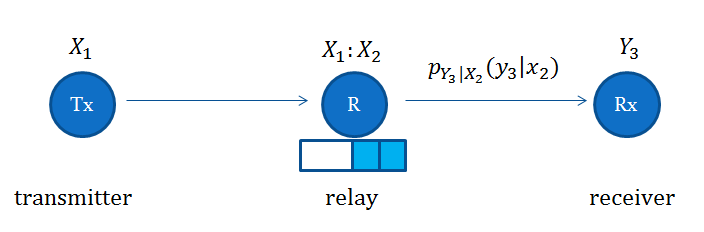}
\caption{Noisy two-hop relay channel with finite battery (Noisy THRC-FB)- noise between the relay and the receiver}
 \label{fig:noisy}
\end{figure}

Now, to insert the noise in to the problem, we consider a binary memoryless channel between the relay and the receiver (the second hop). In this case, noise has the ability to change the transmitted symbol, and of course its energy, randomly.
This means that if channel converts symbol $"1"$ into symbol $"0"$, it would not have energy and if channel converts symbol $"0"$ into symbol $"1"$, it would contain one energy unit. However, this change does not affect the channel state diagram, and thus it is not important in the channel memory.
This system model is illustrated in Fig.\ref{fig:noisy}. All of other considerations are exactly as same as previous model. We call this system as noisy two-hop relay channel with finite battery (Noisy THRC-FB).

Encoding and decoding functions depend on battery size $U$, so we have to include this parameter in our code definition. A $({2^{nR}},n,U)$ code for (Noiseless/Noisy) THRC-FB  consists of a message set $ [1:{{2}^{nR}}]$, an encoder function which maps $m\in [1:{{2}^{nR}}]$ to $x_1^n(m)$, a set of relay encoder functions which map each past received sequence $x_1^{i - 1}$ to ${x_{{2,i}}}(x_1^{i - 1})$ for $i \in [1:n]$, and a decoder function which estimates $\hat{m}$ from the received sequence $y_3^n$ in receiver. We define the average probability of error $P_e^{(n)} = P\{ M \ne \hat M\} $. A rate $R$ is achievable for (Noiseless/Noisy) THRC-FB, if there exists a $({2^{nR}},n,U)$ code such that $\mathop {\lim }\limits_{n \to \infty } P_e^{(n)} = 0$


\section{Noiseless THRC-FB}
In this section we propose an achievable rate for the noiseless THRC-FB.
First we provide a lemma (to be used in the achievability proof), where we state a sufficient condition for existence of steady state probabilities in a finite state Markov chain.
Next, we prove an achievability theorem based on block Markov coding. We used superposition coding for codebook generation for each state. The novel part of our work is in the decoding at the receiver without knowing the state sequence. We use backward decoding and A.E.P Theorem \cite[Theorem 6.6.1]{ash} for message decoding in receiver.
\begin{lemma}\label{lem:steady}
Consider an indecomposable  Markov chain with $r$ possible states. The steady state probabilities exist, if there exists a state $\tilde{S}$ in the state diagram which is accessible from itself in one transition (the probability of returning to itself in next transition is nonzero).
\end{lemma}

%\textit{proof}
\begin{IEEEproof}
It is known that a sufficient condition for existence of the steady state probabilities is that there exist a state $\hat{S}$ and a positive number $n$ such that beginning from any state we can reach state $\hat{S}$ in $n$ steps \cite[Theorem 6.3.2.]{ash}.  Now, we show these conditions hold by choosing $\hat{S}$ to be $\tilde{S}$ and $n$ to be maximum distance between $\tilde{S}$ and any other states in state diagram. Let $S^k$ be the states which has distance $k$ from $\tilde{S}$ and $m$ be the maximum distance, i.e., $m=\max k$. One can reach $\tilde{S}$ from $S^k$ in $l$ steps, where $l$ is an arbitrary integer number that $l\geq k$. Because, after first arrival to $\tilde{S}$, we can stay there (due to the assumption of lemma). Therefore, beginning from any state we can reach $\tilde{S}$ in at least $m$ steps.
\end{IEEEproof}

\begin{theorem}\label{thm:noiseless}
The following rate, $R$, is achievable for Noiseless THRC-FB:
\begin{equation}
R<\underset{p({{x}_{1\left| u \right.}},{{x}_{2\left| u \right.}})}{\mathop{\max }}\,\min \{\sum\limits_{u=0}^{U}{{{\pi }_{u}}H({{X}_{2\left| u \right.}})},\sum\limits_{u=0}^{U}{{{\pi }_{u}}H({{X}_{1\left| u \right.}}\left| {{X}_{2\left| u \right.}} \right.)}\}
\label{thm1}
\end{equation}
where $p(x_{1| u},x_{2| u})$ is chosen such that the state diagram of channel satisfies assumptions of Lemma~\ref{lem:steady}. This means that the state diagram is indecomposable  and there exists a state $\tilde{S}$ in the state diagram which is accessible from itself in one transition. For simplicity, from now on, we call this class of p.m.f.s indecomposable. Note that for $u \in [0:m - 1]$ we must have $p({x_{2\left| u \right.}}) = \left\{ {\begin{array}{*{20}{c}}
{1\begin{array}{*{20}{c}}
{}&{{x_{2\left| u \right.}} = 0}
\end{array}}\\
{0\begin{array}{*{20}{c}}
{}&{{x_{2\left| u \right.}} = 1}
\end{array}}
\end{array}} \right.$.
\end{theorem}

%\textit{proof}
\begin{IEEEproof}
Our scheme uses block Markov coding, where the $B$ blocks of transmissions (each of $n$ symbols) are sent to the relay node to transmit a sequence of $B-1$ independent and identically distributed (i.i.d) messages ${M_b},b \in [1:B - 1]$ while the message of the last block is deterministic. Similarly, the relay node sends $B$ blocks to the receiver in which the message of the first block is deterministic and the messages of the remaining blocks are the same as the transmitter's message with one block delay. At the end of  each block, the relay decodes the message and sends it to the receiver in the next block. Thus, the relay sends a deterministic message in the first block and the transmitter sends a deterministic message in last block. Since the state space is finite, we can control the initial state in each block with at most $U$ transmissions. Thus, we assume that the initial state in each block can be adjusted and for simplicity we do not contain these $U$ transmissions in our further discussions. In fact, by including these transmissions, each block contains $n+U$ bits instead of $n$ bits. %and when we use index $i$ for a bit it means $i'$-th bit in last $n$ bit of block which include our message.

\textit{Codebook generation:}
For each state $u\in\mathcal{U}$, fix an indecomposable p.m.f $p(x_{1| u},x_{2| u})$. Now, for each state $u\in\mathcal{U}$ and for each block $b \in [1:B - 1]$, we generate randomly and independently $2^{nR}$ sequences $x_{2| u }^{n_u+\delta }(m_{b-1})$, where $m_{b-1}\in [1:{{2}^{nR}}]$, each according to $\prod\limits_{i=1}^{n_u+\delta }{p_{X_{2| u }}({x_{2| u,i}})}$.
For each ${m_{b-1}}\in [1:{{2}^{nR}}]$, we generate randomly and conditionally independently ${{K}_{u}}$ sequences $x_{1| u }^{n_u+\delta }(m_{u,b},m_{b-1})$, where $m_{u,b}\in [1:K_u]$, each according to $\prod\limits_{i=1}^{n_u+\delta }{{{p}_{{{X}_{1| u }}| {{X}_{2| u }} }}({{x}_{1| u ,i}}| {{x}_{2| u ,i}}(m_{b-1}) )}$, where $K_u$ is the size of the transmitter's codebook of state $u$ and $m_{u,b}$ is the message of the transmitter's codebook of state $u$ in block $b$. We have ${m_0} = {m_B} = 1$.
{In fact, we do a kind of rate splitting in transmitter in which $K_u=2^{nR_u}$ and $R_u$ is the rate of each codebook.
}
 The codewords are shown in Fig.~\ref{fig:coding}.

In addition, we generate ${{2}^{nR}}$ i.i.d random variables $U(m_{b-1})$, where ${m_{b-1}}\in [1:{{2}^{nR}}]$, with p.m.f $\pi_u$. These random variables will be used as initial state of channel in block $b$.

We remark that only the first $n_u$ bits (in each codeword) contain message and we find $n_u$ for each state such that the error probability tends to zero. Other $\delta$ bits are generated to protect channel's statistical properties (Markovity) from change. This means that we generate $\delta$ joint random bits from the p.m.f $p(x_{1| u},x_{2| u})$ for each message set {$(m_{u,b},m_{b-1})$}. Thus, if $n_u$ bits of codeword of state $u$ are sent completely before the codewords of other states, sending these $\delta$ bits would prevent the changing of statistical properties of channel state diagram. $\delta$ can be chosen as large as $n-min(n_u)$ to satisfy the above condition. The transmission strategy is described in the following.

\begin{figure}
\centering
\includegraphics[width=6cm]{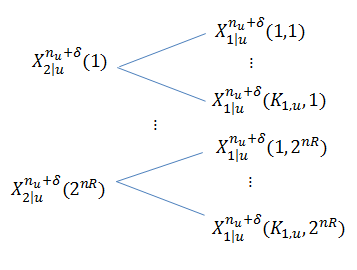}
\caption{Superposition coding for each state with joint p.m.f $p({{x}_{1\left| u \right.}},{{x}_{2\left| u \right.}})$}
 \label{fig:coding}
\end{figure}

\textit{Encoding (at the beginning of block $b$)}

\textbf{Transmitter}:
To send message ${m_b}\in [1,\prod\limits_{u=0}^{U}{{{K}_{u}}}]$ in block $b$, transmitter maps the message into a message vector $[{{m}_{0,b}},{{m}_{1,b}},...,{{m}_{U,b}}]$, ${{m}_{u,b}}\in [1:{{K}_{u}}]$. Then, it selects the codeword $x_{1| u }^{n_u+\delta }(m_{u,b},m_{b-1})$ from the codebook corresponding to state $u$. In addition, we set a vector as: $[{l_{0,b}} = 1,{l_{1,b}} = 1,...,{l_{U,b}} = 1]$. For encoding in block $b$, starting from the beginning of the block with $u_1$ as initial state, we send ${x_{1\left| {u_1,{l_{u_1,b}}} \right.}}({m_{u_1,b}},{m_{b - 1}})$ and we set ${l_{u_1,b}} = {l_{u_1,b}} + 1$. We repeat this procedure for next transmissions. Encoding procedure is shown completely in Fig \ref{fig:flowchart}. This procedure is similar to the encoding introduced in \cite{popovski}.

\begin{figure}
\centering
\includegraphics[width=5.7cm]{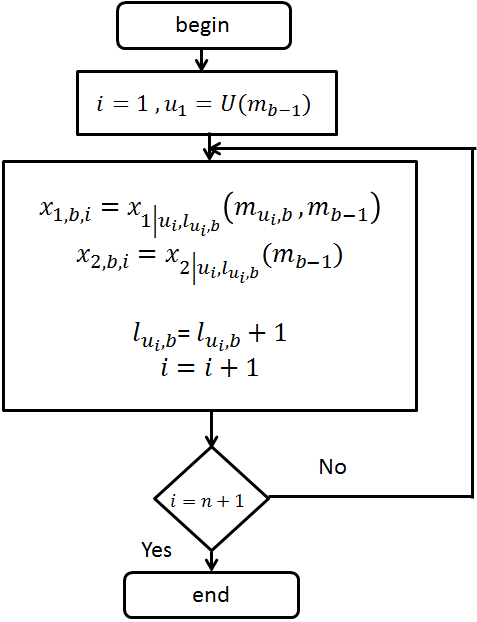}
\caption{Flowchart of encoding block $b$ in transmitter and relay}
 \label{fig:flowchart}
\end{figure}

\textbf{Relay:}
The relay sends message $m_{b-1}$ in block $b$, so it selects codeword $m_{b-1}$ from each codebook as the message of that codebook and chooses $U(m_{b-1})$ as initial state. The next stages are the same as the encoding at the transmitter and are shown in Fig. \ref{fig:flowchart}.

\textit{Decoding (at the end of block $b$)}

\textbf{Relay:}
The state sequence in block $b$, $u_b^n$, is known at the relay, because it knows its battery storage. Thus, for each $u \in [0:U]$, the relay makes a set ${A_u} = \{ i\left| {{u_i} = u} \right.\}$. If  $\left| {{A_u}} \right| \ge {n_u}$, the relay looks for an ${{\hat m}_{u,b}} = \{ {m_{u,b}}\left| {{x_{2\left| {u,k} \right.}}({m_{u,b}}) = {x_{2,b,{i_k}}}} \right.\} $, $k \in [1:{n_u}]$ and ${i_1} \le {i_2} \le ... \le {i_{{n_u}}}$. This procedure is shown in Fig. \ref{fig:flowchartdecode}.
In the error probability analysis, we find the conditions which guarantee the ${{\hat m}_{u,b}}$ to be unique. Then, the relay forms the vector $[{{\hat m}_{0,b}},{{\hat m}_{1,b}},...,{{\hat m}_{U,b}}]$  and by the inverse of mapping used in encoding, it can decode $\hat{m}_b$. This procedure is like the decoding introduced in \cite{popovski}.

\begin{figure}
\centering
\includegraphics[width=8cm]{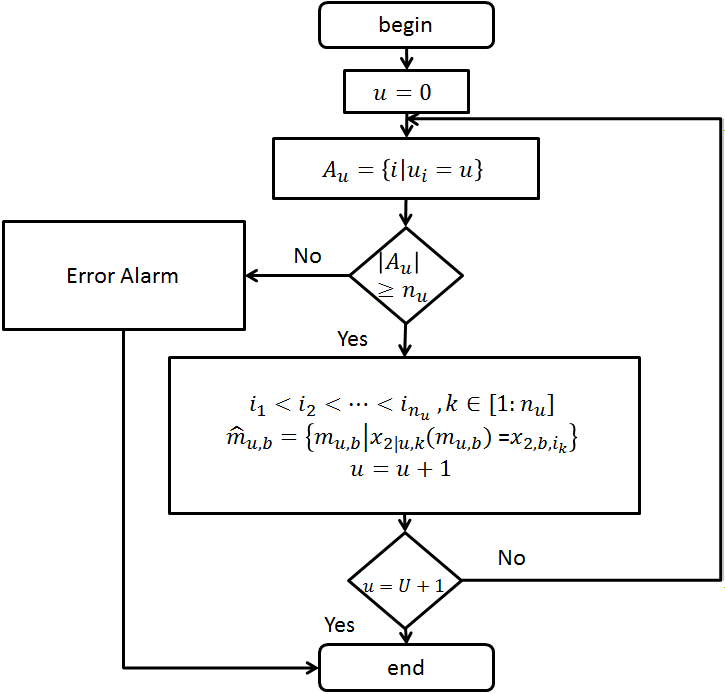}
\caption{Flowchart of decoding block $b$ in relay}
 \label{fig:flowchartdecode}
\end{figure}

\textbf{Receiver}:
The receiver uses backward decoding. In the last block, the transmitted message is fixed, i.e., $m_B=1$. Assume that the relay transmits the message $m'_{B-1}$ in last block, where $m'_{B-1}\in [1:{{2}^{nR}}]$. The receiver runs the flowchart shown in Fig. \ref{fig:flowchart} for each $m'_{B-1}$ and computes the sequence generated at the relay for each $m'_{B-1}$. We call these sequences $X_2^n(m_B = 1,m'_{B - 1})$.%, this notation is because of that $X_2^n$ is a function of transmitter and relay message in each block and in last block $B$ we have $m_B=1$.
Indeed, $X_2^n(m_B = 1,m'_{B - 1})$ is the sequence which is received at the receiver if the relay sends message $m'_{B-1}$ in last block.
To perform the above decoding, the receiver does not need the state sequence, because: 1)the codebooks and initial states are shared; 2)there is not any noise; 3)as we see in the flowchart of Fig. \ref{fig:flowchart}, since the transmitter's message (in last block $m_B=1$), relay's message ($m'_{B-1}$) and the initial state are known, after every transmission the next state can be determined and $X_2^n(m_B = 1,m'_{B - 1})$ could be derived for each $m'_{B - 1}\in [1:{{2}^{nR}}]$.
Then, the receiver looks for a unique ${\hat{m}}_{B-1}$, such that its computed sequence is equal to the received sequence. We show that $2^{nR}$ probable received sequences are not equal with probability 1. When message ${m}_{B-1}$ is decoded, the transmitter's message in block $B-1$ is known and the above procedure can be repeated to decode the previous blocks messages.

\textit{Error probability analysis:}
The probability of error is upper bounded by the sum of probabilities of error in the relay and the receiver.
The error events at the relay in each block are:
\begin{itemize}
\item  ${\varepsilon ^{(1)}}=$ Relay does not receive the codeword of at least one of the codebooks (corresponds to a state $u$) completely. We define $\varepsilon _u^{(1)}$ as the event in which the relay does not receive the codeword of codebook $u$ completely.
\item  ${\varepsilon ^{(2)}}=$ There are more than one equal codewords with received sequence in at least one of codebooks. We define $\varepsilon _u^{(2)}$ as the event in which there are more than one equal codewords with the received sequence of codebook $u$.
% {\color{red} after writing the typicality with formulas (above), also write this even in math form}
\end{itemize}
where it is seen that $P({\varepsilon ^{(i)}}) \le \sum\limits_u {P(\varepsilon _u^{(i)})} ,i \in \{ 1,2\}$.

First, we consider the ${\varepsilon ^{(1)}}$. Recall that the state sequence has steady state by Lemma~\ref{lem:steady}. In addition, based on a result in \cite[Theorem 6.6.3.]{ash}, in a finite Markov chain with steady state, the relative frequency of being in a state $u$ converges to the steady state probability ${\pi _u}$, in probability. Thus, if we choose ${n_u} = n({\pi _u} - \epsilon )$, the event ${\varepsilon ^{(1)}}$ does not occur with probability 1 and $P({\varepsilon ^{(1)}})$ goes to zero for large enough $n$.

Based on  Lemma (\ref{lemmaaepsilon}), the probability of the second error event (${\varepsilon ^{(2)}}$) goes to zero if:
\begin{equation}\label{eqn:EE2}
\frac{{\log ({K_{u}})}}{{{n_u}}} < H({X_{1\left| u \right.}}\left| {{X_{2\left| u \right.}}} \right.) - \epsilon \
\end{equation}

{
\begin{lemma}
\label{lemmaaepsilon}
Fix a joint p.m.f $p(u,x)$ and generate a random sequence $U^n$ according to $\prod\limits_{i = 1}^n {{p_U}({u_i})} $, then generate randomly and conditionally independently $2^{nR}$ sequences ${X^n}(m),m \in [1:{2^{nR}}]$, each according to $\prod\limits_{i = 1}^n {{p_{X\left| U \right.}}(\left. {{x_i}} \right|{u_i})}$. If we have $R < H(X\left| U \right.)$, then probability of  the event $\{{X^n}(m) = {X^n}(m'),m \ne m'\}$, would tend to zero.
\end{lemma}

\begin{IEEEproof}
Based on packing lemma \cite{elGamal}, if the sequence $X^n(1)$ is passed through a discrete memoryless channel $\prod\limits_{i = 1}^n {{p_{Y\left| X \right.}}(\left. {{y_i}} \right|{x_i})} $ and the sequence $Y^n$ is constructed and if $R < I(X;Y\left| U \right.)$, then we would have $P(({U^n},{X^n}(m),{Y^n}) \in A_\varepsilon ^n(p(u,x,y))) \to 0,m \ne 1$. Now we consider a $p(y\left| x \right.)$  such  that $p(y = x\left| x \right.) = 1,p(y \ne x\left| x \right.) = 0$, so we obtain $Y^n=X^n(1)$. Now we determine $A_\varepsilon ^n(p(u,x,y))$. By definition we have:
\begin{equation*}
\scalebox{0.86}[1]{$A_\varepsilon ^n(p(u,x,y)) = \{ ({x^n},{u^n},{y^n})\left| {\left| {\pi (x,u,y\left| {{x^n},{u^n},{y^n}} \right.) - p(u,x,y)} \right|} \right. \le$}  
\end{equation*}
\begin{equation}
\varepsilon p(u,x,y),\forall u \in \mathcal{U},\forall x \in \mathcal{X},\forall y \in \mathcal{Y}\}
\label{Aeps}
\end{equation}
in which ${\pi (x,u,y\left| {{x^n},{u^n},{y^n}} \right.)}$ is the percentage of repetition of $u,x,y$ in the sequence ${x^n},{u^n},{y^n}$. 

For conditional distribution described above, we have $p\left( {u,x,y \ne x} \right) = 0,p\left( {u,x,y = x} \right) = p\left( {u,x} \right)p\left( {y\left| x \right.} \right) = p\left( {u,x} \right)$, and thus:
\begin{equation}
\pi \left( {u,x,y \ne x\left| {{u^n},{x^n},{y^n}} \right.} \right) = 0
\label{pi0}
\end{equation}
\begin{equation}
\pi \left( {u,x,y = x\left| {{u^n},{x^n},{y^n}} \right.} \right) = \pi \left( {u,x\left| {{u^n},{x^n}} \right.} \right)
\label{pi1}
\end{equation}

By inserting (\ref{pi0}) and (\ref{pi1}) in (\ref{Aeps}), we get:
\begin{equation}
\label{Aepsilon}
\scalebox{0.89}[1]{$A_\varepsilon ^n\left( {p\left( {u,x,y} \right)} \right) = \left\{ {\left( {{u^n},{x^n},{y^n}} \right)\left| {\left( {{u^n},{x^n}} \right) \in A_\varepsilon ^n\left( {p\left( {u,x} \right)} \right),} \right.{x^n} = {y^n}} \right\}$}
\end{equation}
which results in:
\begin{equation*}
P(({U^n},{X^n}(m),{Y^n}) \in A_\varepsilon ^n(p(u,x,y)))
\end{equation*}
\begin{equation*}
= P(({U^n},{X^n}(m)) \in A_\varepsilon ^n(p(u,x)) \cap {X^n}(m) = {X^n}(1))
\end{equation*}

So by packing lemma \cite{elGamal} for $m \ne 1$, we have:
\begin{equation*}
P(({U^n},{X^n}(m),{Y^n}) \in A_\varepsilon ^n(p(u,x,y))) < \varepsilon '
\end{equation*}
\begin{equation*}
P(({U^n},{X^n}(m)) \in A_\varepsilon ^n(p(u,x)) \cap {X^n}(m) = {X^n}(1)) < \varepsilon '
\end{equation*}
And by complementing above equation we have:
\begin{equation}
 \scalebox{0.9}[1]{$P(({U^n},{X^n}(m)) \notin A_\varepsilon ^n(p(u,x)) \cup {X^n}(m) \ne {X^n}(1)) > 1 - \varepsilon '$}
\label{complement}
\end{equation}
we apply union bound to obtain:
\begin{equation*}
P(({U^n},{X^n}(m)) \notin A_\varepsilon ^n(p(u,x)) \cup {X^n}(m) \ne {X^n}(1))
\end{equation*}
\begin{equation}
 \scalebox{0.9}[1]{$\le P(({U^n},{X^n}(m)) \notin A_\varepsilon ^n(p(u,x))) + P({X^n}(m) \ne {X^n}(1))$}
\label{pAeps}
\end{equation}

In addition, by Joint A.E.P Theorem \cite[Theorem 7.6.1]{cover}, we have:
\begin{equation}
P(({U^n},{X^n}(m)) \notin A_\varepsilon ^n(p(u,x))) < \varepsilon ''
\label{jointAeps}
\end{equation}
By combining (\ref{complement})-(\ref{jointAeps})  we obtain:
\begin{equation*}
P({X^n}(m) \ne {X^n}(1)) > 1 - \varepsilon ' - \varepsilon '' = 1 - \varepsilon '''
\end{equation*}

So by packing lemma \cite{elGamal} and discussions given in the first paragraph of proof and the fact $X=Y$, if we have:
\begin{equation*}
R < I\left( {X;Y\left| U \right.} \right)=I\left( {X;X\left| U \right.} \right) = H\left( {X\left| U \right.} \right)
\end{equation*}
then we can derive inequality (\ref{inequal}):
\begin{equation}
\label{inequal}
P({X^n}(m) = {X^n}(1)) < \varepsilon ''' , m \ne 1
\end{equation} 

Where $\varepsilon '''$ could take every small value for large enough $n$.
Repeating this argument for each ${X^n}(m), m \in [1:{2^{nR}}]$, completes the proof.
\end{IEEEproof}
 }

If we substitute ${n_u} = n({\pi _u} - \epsilon )$ in \eqref{eqn:EE2}, we obtain:
{
\begin{equation}
{n_u} = n\left( {{\pi _u} - \epsilon } \right) \to {n_u} < n\left( {{\pi _u}} \right) \to \frac{1}{{{n_u}}} > \frac{1}{{n\left( {{\pi _u}} \right)}}
\label{eq29}
\end{equation}
\begin{equation*}
R = \frac{{{{\log }_2}\mathop \prod \nolimits_{u = 0}^U {K_u}}}{n} = \mathop \sum \limits_{u = 0}^U \frac{{{\pi _u}{{\log }_2}{K_u}}}{{n{\pi _u}}}
\end{equation*}
\begin{equation}
< \mathop \sum \limits_{u = 0}^U \frac{{{\pi _u}{{\log }_2}{K_u}}}{{{n_u}}} 
\label{eq30}
\end{equation}
}
\begin{equation}
\label{eq31}
R < \sum\limits_{u = 0}^U {{\pi _u}H({X_{1\left| u \right.}}\left| {{X_{2\left| u \right.}}} \right.)}
\end{equation}
{where equation (\ref{eq30}) is derived by equation (\ref{eq29}) and equation (\ref{eq31}) is derived by equations (\ref{eqn:EE2}) and (\ref{eq30}).}

For error probability analysis in receiver, recall that in the last block, the transmitter's message is deterministic and so the receiver derives $X_2^n(m_B = 1,m'_{B - 1})$, ${{{m}^{'}}_{B-1}}\in [1:{{2}^{nR}}]$. Since we assume that relay sends $X_2^n(m_B = 1,m_{B - 1}=1)$,
If at least one $X_2^n({m_B} = 1,m'_{B - 1}\ne1)$ becomes equal to $X_2^n(m_B = 1,m_{B - 1}=1)$, the error occurs in the receiver.

%{\color{red} better to put this paragraph in a lemma}

\begin{lemma}
For each $m'_{B - 1}\in [1:{{2}^{nR}}]$, the $X_2^n(m_B = 1,m'_{B - 1})$ is an independent regular Markov source.
\end{lemma}

%\textit{proof}
\begin{IEEEproof}
Independency can be easily deduced from the codebook generation because the initial states and all codewords in each codebook are generated independently. To show that these sequences are regular markov sources, we define a new Markov chain as: denoting the state sequence as $U_1,U_2,U_3,\ldots$, the new Markov chain is $S_1=(U_1,U_2),S_2=(U_2,U_3),S_3=(U_3,U_4),\ldots$, as we described in section (II). Since ${X_2}_i$ is determined by $U_i,U_{i+1}$, we have ${X_{2,i}}=f(S_i)$, where $f$ is a deterministic function. Thus, ${X_{2,i}}$ is a markov source. Moreover, the assumptions of Lemma~\ref{lem:steady} are also satisfied by the new Markov chain $S_i$ and so steady state probabilities exist. This shows that ${X_{2,i}}$ is a regular Markov source.
\end{IEEEproof}

Since a regular Markov source is ergodic \cite[Theorem 6.6.2]{ash}, ${X_{2,i}}$ satisfies conditions of Asymptotic EquiPartition Property (A.E.P) Theorem \cite[Theorem 6.6.1]{ash} which states that if we have $2^{nR}$ i.i.d Markov sources with entropy rate $H\{\underline{X}\}$, these Markov sources are not equal with probability 1 when $R<H\{\underline{X}\}$.
Now, we derive the entropy rate of ${X_{2,i}}$. Since ${X_{2,i}}$ is not unifilar, we cannot use entropy rate of unifilar Markov sources {(a unifilar Markov source is a Markov source in which the present state $U_n$ and the present output $X_n$ compute the next state $U_{n+1}$)}. We use the followings (for simplicity, the index $B$ for block number is omitted from the equations):
\begin{equation}
\scalebox{0.9}[1]{$H({{X}_{{{2},{n}}}}\left| {{X}_{{{2},{n-1}}}},...,{{X}_{{{2},{1}}}} \right.)\ge H({{X}_{{{2},{n}}}}\left| {{U}_{n}},{{X}_{{{2},{n-1}}}},...,{{X}_{{{2},{1}}}} \right.)$}
\label{eq6}
\end{equation}
\begin{equation}
=H({{X}_{{{2},{n}}}}\left| {{U}_{n}} \right.)=H({{X}_{{{2},{1}}}}\left| {{U}_{1}} \right.)=\sum\limits_{u=0}^{U}{{{\pi }_{u}}H({{X}_{2\left| u \right.}})}
\label{eq7}
\end{equation}
%{\color{red} tell the reason for each inequality and equality}
The inequality (\ref{eq6}) follows from the fact that conditioning does not increase the entropy and the equation (\ref{eq7}) holds since conditioning on $U_n$, the distribution of $X_{2,n}$ is determined independent of ${{X}_{{{2},{n-1}}}},...,{{X}_{{{2},{1}}}}$ and our processes are stationary.
Therefore, if $R<\sum\limits_{u=0}^{U}{{{\pi }_{u}}H({{X}_{2\left| u \right.}})}$, then the generated sequences are not equal with probability 1 by A.E.P Theorem. This completes the proof.
\end{IEEEproof}

{Note that to find $K_u$, first we have to solve optimization problem in equation (\ref{thm1}), so we can determine ${p({x_{1\left| u \right.}},{x_{2\left| u \right.}})}$. Next, we calculate $H\left( {{X_{1\left| u \right.}}\left| {{X_{2\left| u \right.}}} \right.} \right)$ and we consider ${K_u} < H\left( {{X_{1\left| u \right.}}\left| {{X_{2\left| u \right.}}} \right.} \right)$.

}
\section{Noisy THRC-FB}

\begin{theorem}\label{thm:noisy}
The following rate, $R$, is achievable for Noisy THRC-FB:
\begin{equation}
\scalebox{0.9}[1]{$R < \mathop {\max }\limits_{p({x_{1\left| u \right.}},{x_{2\left| u \right.}})} \min \{ \sum\limits_{u = 0}^U {{\pi _u}I({X_{2\left| u \right.}};{Y_{3\left| u \right.}})} ,\sum\limits_{u = 0}^U {{\pi _u}H({X_{1\left| u \right.}}\left| {{X_{2\left| u \right.}}} \right.)} \}$}
\end{equation}
where $P({{x}_{1\left| u \right.}},{{x}_{2\left| u \right.}})$ is an indecomposable p.m.f and for $u \in [0:m - 1]$ we must have $p({x_{2\left| u \right.}}) = \left\{ {\begin{array}{*{20}{c}}
{1\begin{array}{*{20}{c}}
{}&{{x_{2\left| u \right.}} = 0}
\end{array}}\\
{0\begin{array}{*{20}{c}}
{}&{{x_{2\left| u \right.}} = 1}
\end{array}}
\end{array}} \right.$.
\end{theorem}

%\textit{proof}
\begin{IEEEproof}
All proof steps of this theorem are exactly the same as the proof of Theorem~\ref{thm:noiseless}, except the decoding at the receiver and the error probability analysis in receiver. Thus, we only highlight the differences.

%{\color{red} apply the comment I gave you in the previous section here too}
\textit{Receiver decoding:}
Receiver uses backward decoding. As we showed for noiseless THRC-FB, the relay computes $X_2^n(m_B = 1,m'_{B - 1})$ for each $m'_{B - 1}\in [1:{{2}^{nR}}]$. Then, the receiver looks for ${\hat{m}}_{B-1}$ which satisfies $(X_2^n(m_B = 1,{{\hat m}_{B - 1}}),Y_{3,B}^n) \in \tau _\varepsilon ^{(n)}$, where the $Y_{3,B}^n$ is received sequence in receiver in block $B$.

\textit{Error probability analysis in receiver:}
As we showed in { Lemma 2}, $X_2^n(m_B = 1,m'_{B - 1})$ are independent and identical Markov sources (for  ${{m}^{'}}_{B-1}\in [1:{{2}^{nR}}]$). If the relay sends the sequence $X_2^n(m_B= 1,m_{B - 1}=1)$ in block $B$, the received sequence in the receiver becomes independent of the other sequences $X_2^n({m_B} = 1,m'_{B - 1}\ne1)$. Thus, we calculate probability of the event in which $(X_2^n({m_B} = 1,{{\hat m}_{B - 1}} \ne 1),Y_{3,B}^n) \in \tau _\varepsilon ^{(n)}$ (for simplicity, the index $B$ for block number is omitted from equations):
\begin{align}
{{P}_{e}}&=\sum{\underset{( \underline{\alpha },\underline{\beta } )\in A_{\varepsilon }^{n}}{\mathop \sum }\,P( X_{2}^{n}=\underline{\alpha } )P( Y_{3}^{n}=\underline{\beta } )}\nonumber\\
&\le {{2}^{-n\left( H\left\{ \underline{{{X}_{2}}} \right\}+\varepsilon  \right)}}{{2}^{-n\left( H\left\{ \underline{{{Y}_{3}}} \right\}+\varepsilon  \right)}}{{2}^{n\left( H\left\{ \underline{{{X}_{2}}},\underline{{{Y}_{3}}} \right\}-\varepsilon  \right)}}
\end{align}

So the upper bound on the error probability is:
\begin{equation}
{{P}_{error}}\le {{2}^{nR}}{{2}^{-n\left( H\left\{ \underline{{{X}_{2}}} \right\}+H\left\{ \underline{{{Y}_{3}}} \right\}-H\left\{ \underline{{{X}_{2}}},\underline{{{Y}_{3}}} \right\}-3\varepsilon  \right)}}
\label{eq10}
\end{equation}

Thus the error probability tends to zero, if:
\begin{equation}
\scalebox{0.85}[1]{$R<H\left\{ \underline{{{X}_{2}}} \right\}+H\left\{ \underline{{{Y}_{3}}} \right\}-H\left\{ \underline{{{X}_{2}}},\underline{{{Y}_{3}}} \right\}=\underset{n\to \infty }{\mathop{\lim }}\,\frac{1}{n}\left( I\left( X_{2}^{n};Y_{3}^{n} \right) \right)$}
\label{eq11}
\end{equation}

In addition, ${Y_3^n}$ is stationary, so we have:
\begin{equation*}
\mathop {\lim }\limits_{n \to \infty } \frac{1}{n}H\left( {Y_3^n} \right) = \mathop {\lim }\limits_{n \to \infty } H({Y_{{3,n}}}\left| {{Y_{{3,{n - 1}}}},{Y_{{3,{n - 2}}}},...,{Y_{{3,1}}})} \right.
\end{equation*}

The term in the left hand side can be written as:
\begin{equation}
\scalebox{0.9}[1]{$H({Y_{{3,n}}}\left| {{Y_{{3,{n - 1}}}},...,{Y_{{3,1}}})} \right. \ge H({Y_{{3,n}}}\left| {{U_n},{Y_{{3,{n - 1}}}},...,{Y_{{3,1}}})} \right.$}
\label{eq12}
\end{equation}
\begin{equation}
= H({Y_{{3,n}}}\left| {{U_n}} \right.) = H({Y_{{3,1}}}\left| {{U_1}} \right.) = \sum\limits_{u = 0}^U {{\pi _u}H({Y_{3\left| u \right.}})}
\label{eq13}
\end{equation}
%{\color{red} tell the reason for each inequality and equality}
The reasons for (\ref{eq12}) and (\ref{eq13}) are exactly the same as the ones for (\ref{eq6}) and (\ref{eq7}).
On the other hand, we have a binary memoryless channel which satisfies:
\begin{equation*}
H\left( {Y_3^n\left| {X_2^n} \right.} \right) = \mathop \sum \limits_{k = 1}^n H\left( {{Y_{{3,k}}}\left| {{X_{{2,k}}}} \right.} \right)
\end{equation*}
which can be continued as:
\begin{align}
H\left( {{Y_{{3,k}}}\left| {{X_{{2,k}}}} \right.} \right)& = \mathop \sum \limits_x H\left( {{Y_{{3,k}}}\left| {{X_{{2,k}}} = x} \right.} \right)p\left( {{X_{{2,k}}} = x} \right)\nonumber\\
 &= \mathop \sum \limits_x H\left( {{Y_{{3,k}}}\left| {{X_{{2,k}}} = x} \right.} \right)\mathop \sum \limits_u {\pi _u}p\left( {{X_{{2,k}}} = x\left| u \right.} \right)
\label{eq14}\\
 &= \mathop \sum \limits_u {\pi _u}\mathop \sum \limits_x H\left( {{Y_3}\left| {{X_2} = x} \right.} \right)p\left( {{X_2} = x\left| u \right.} \right)
\label{e15}\\
& = \mathop \sum \limits_u {\pi _u}H\left( {{Y_{3\left| u \right.}}\left| {{X_{2\left| u \right.}}} \right.} \right)
\label{eq16}
\end{align}
where (\ref{eq14}) is due to the law of total probability and the equation (\ref{e15}) holds thanks to the stationarity of ${{X_{{2,k}}},{Y_{{3,k}}}}$. By combining (\ref{eq11}), (\ref{eq13}) and (\ref{eq16}), we derive:
\begin{equation}
\frac{1}{n}\left( {I\left( {X_2^n;Y_3^n} \right)} \right) \ge \mathop \sum \limits_u {\pi _u}I\left( {{Y_{3\left| u \right.}};{X_{2\left| u \right.}}} \right)
\label{eq17}
\end{equation}

Now, based on (\ref{eq10}) and (\ref{eq17}), we see that if  $R < \mathop \sum \limits_u {\pi _u}I\left( {{Y_{3\left| u \right.}};{X_{2\left| u \right.}}} \right)$, then the error probability in receiver tends to zero. Hence proof is complete.
\end{IEEEproof}

\section{Discussions and Conclusions}
We studied a two-hop channel with an RF energy harvesting relay (with finite battery size), where the transmitter jointly transfers information and energy to the relay. Modeling the energy level at the relay's battery with states, we propose the achievability schemes for the channel with memory, where the main challenge was the unknown state at the receiver. Our proposed schemes work for the noiseless channel and the channel with noisy second hop.

\textbf{Noisy first hop}: By considering the noise in the first hop (between the transmitter and relay), the transmitter does not know the relay's battery level. Thus, the state is not available to the transmitter and as a result the receiver cannot compute the possible transmitted sequences of relay (for each message). Therefore, the proposed schemes are not readily extended to this case. Designing appropriate coding schemes for this channel is our ongoing research work.

\textbf{Upper bound}: Due to the channel memory, the problem of finding a tight outer bound for this system model cannot be tackled by using standard inequalities used in converse proofs.
%\begin{itemize}
%\item
%\item
%%\item[3)]  Extend these results to a general channel with memory in order to find a single letter capacity.
%
%\end{itemize}

\end{document}